\documentclass[twocolumn,showpacs,superscriptaddress,nofootinbib,floatfix,10pt]{revtex4-1}

\usepackage[utf8]{inputenc}
\usepackage{float}
\usepackage{amsmath,amssymb,amsfonts,bm}
\usepackage{epstopdf}
\usepackage{graphicx}
\usepackage{multirow}
\usepackage[usenames,dvipsnames]{color}
\usepackage[
colorlinks=true,
linkcolor=blue,
breaklinks=true,
urlcolor=blue,
citecolor=blue]{hyperref}

\newcommand{\B}{}
\newcommand{\Tr}[1]{{{\rm Tr}\!\left[#1\right]}}

\newcommand{\D}[0]{\mathrm{d}}

\newcommand{\co}[0]{{(Color online) }}
\newcommand{\be}{\begin{equation}}
\newcommand{\ee}{\end{equation}}
\renewcommand{\vec}{\bm}

\graphicspath{{figs.dir/}}

\newcommand{\rub}{\affiliation{Institut f\"ur Theoretische Physik II, Ruhr-Universit\"at Bochum,  Bochum D-44780, Germany }}

\newcommand{\fzj}{\affiliation{Institute for Advanced Simulation, Institut f\"ur Kernphysik and J\"ulich Center for Hadron Physics, Forschungszentrum J\"ulich,  J\"ulich D-52425, Germany}}

\newcommand{\itp}{\affiliation{CAS Key Laboratory of Theoretical Physics, Institute of Theoretical Physics, Chinese Academy of Sciences, Beijing 100190, China}}

\newcommand{\ucas}{\affiliation{School of Physical Sciences, University of Chinese Academy of Sciences, Beijing 100049, China}}

\newcommand{\itep}{\affiliation{Institute for Theoretical and Experimental Physics NRC ``Kurchatov Institute'', Moscow 117218, Russia }}

\newcommand{\lebedev}{\affiliation{P.N. Lebedev Physical Institute of the Russian Academy of Sciences, Moscow 119991, Russia}}

\newcommand{\mipt}{\affiliation{Moscow Institute of Physics and Technology,  Moscow 141700, Russia}}
\newcommand{\znu}{\affiliation{School of Physics, Central South University, Changsha 410083, China}}
\begin{document}

\title{Is the existence of a $J/\psi J/\psi$ bound state plausible?}
\author{Xiang-Kun Dong}\email{dongxiangkun@itp.ac.cn}
\itp\ucas

\author{Vadim Baru}\email{vadim.baru@tp2.rub.de}
\rub \itep

\author{Feng-Kun Guo}\email{fkguo@itp.ac.cn}
\itp \ucas

\author{Christoph~Hanhart}\email{c.hanhart@fz-juelich.de}
\fzj

\author{Alexey Nefediev}\email{nefediev@lebedev.ru}
\lebedev
\mipt

\author{Bing-Song Zou} 
\itp\ucas\znu

\begin{abstract}
 In a recent measurement LHCb reported pronounced structures in the $J/\psi J/\psi$ spectrum. One of the various possible explanations of those is that they emerge from non-perturbative interactions of vector charmonia. It is thus important to understand whether it is possible to form a bound state of two charmonia
 interacting through the exchange of gluons, which hadronise into two pions at the longest distance. In this paper, we demonstrate that, given our current understanding of hadron-hadron interactions, the exchange of correlated light mesons (pions and kaons) is able to provide sizeable attraction to the di-$J/\psi$ system, and it is possible for two $J/\psi$ mesons to form a bound state.
 As a side result we find from an analysis of the data for the $\psi(2S)\to J/\psi \pi\pi$ transition including both $\pi\pi$ and $K\bar K$ final state interactions an improved value for the $\psi(2S)\to J/\psi$ transition chromo-electric polarisability:
 $|\alpha_{\psi(2S)J/\psi}|= (1.8\pm 0.1)~\mbox{GeV}^{-3}$,
 where the uncertainty also includes the one induced by the final state interactions.
\end{abstract}

\maketitle

\section{Introduction}

Recently the LHCb Collaboration announced the observation of resonance-like structures in the double-$J/\psi$ invariant mass distribution~\cite{Aaij:2020fnh}. In particular, there is a narrow structure at about 6905~MeV with a width of around 80~MeV,
which is a promising candidate for a fully-charmed tetraquark $cc\bar c\bar c$ state hypothesised to exist long time ago \cite{Iwasaki:1975pv,Chao:1980dv,Badalian:1985es}, see also Refs.~\cite{Maiani:2020pur,Chao:2020dml,Richard:2020hdw} for comments on this newly observed resonant structure. 
The LHCb discovery has already spurred a series of model studies of the fully-charmed tetraquarks~\cite{Bedolla:2019zwg,Deng:2020iqw,Yang:2020rih,Wang:2020ols,Jin:2020jfc,Lu:2020cns,Becchi:2020uvq,Sonnenschein:2020nwn,Giron:2020wpx,Karliner:2020dta,Wang:2020dlo,Zhao:2020nwy,Faustov:2020qfm,Zhang:2020xtb,Zhu:2020xni,Yang:2020wkh,Zhao:2020zjh,Faustov:2021hjs,Ke:2021iyh,Yang:2021hrb,Mutuk:2021hmi,Li:2021ygk,Wang:2021kfv,Majarshin:2021hex,Pal:2021gkr,Tiwari:2021tmz,Nefediev:2021pww}. Indeed, some tetraquark models predict a state with the mass close to the LHCb Breit-Wigner mass and, in addition, the partial width of the inclusive decays of a fully-charm tetraquark state into final states containing a pair of charmed hadrons is indeed expected to be of the order 100~MeV~\cite{Chao:1980dv,Anwar:2017toa,Becchi:2020uvq}.

Meanwhile, one should keep in mind that the line shape of a resonance may deviate from the Breit-Wigner parametrisation significantly if it is located near the threshold of another channel that it can couple with. Several works have investigated the impacts of such threshold effects on the pole location of the $X(6900)$~\cite{Wang:2020wrp,Dong:2020nwy,Cao:2020gul,Liang:2021fzr}. Furthermore, the coupled-channel analysis of the LHCb data performed in Ref.~\cite{Dong:2020nwy} has shown that the structures observed in the double-$J/\psi$ system can be understood as being closely linked to non-perturbative scattering of vector charmonia, leading to hints of the existence of a near-threshold state in the $J/\psi J/\psi$ channel, denoted as $X(6200)$. This finding was confirmed independently in Ref.~\cite{Liang:2021fzr}. Clearly, this observation should be followed by a study on if the existence of a double-$J/\psi$ bound state is consistent with our present understanding of hadron-hadron interactions.  The interaction of two heavy quarkonia, considered as color dipoles in the nonperturbative quantum chromodynamics (QCD) vacuum, was studied in Ref.~\cite{Fujii:1999xn} with the help of the low-energy QCD theorems while the modification of the spectrum of such quarkonia due to nonperturbative gluons was addressed in Ref.~\cite{Leutwyler:1980tn}. In Ref.~\cite{Brambilla:2015rqa}, the interaction between a pair of bottomonia was studied, which at the hadronic level occurs through the exchange of two pions. The same interaction is expected to be operative in the charm sector, too. Indeed, the interactions between a pair of charmonia should be of the same order as of bottomonia, given that they are provided by the exchange of soft gluons, which hadronise into exchanges of pions and heavier hadrons at large distances. {More generally, the interactions between different pairs of fully-heavy hadrons should be similar, at least qualitatively. In this context it is important to notice that a recent lattice study showed that the strong interaction between a pair of $\Omega_{ccc}$ baryons is strong enough to produce a near-threshold bound state~\cite{Lyu:2021qsh}.} It is, therefore, important and timely to investigate such an interaction between a pair of charmonia to see if its strength is of the right order of magnitude to generate a near-threshold pole in the double-$J/\psi$ system, which could be interpreted as a molecular $X(6200)$ state.

Since the scalar-isosclar two-pion and two-kaon interactions are known to be non-perturbative, giving rise to the generation
of scalar resonances, namely the $f_0(500)$ and the $f_0(980)$, we include the $\pi\pi$ as well as
the $\bar KK$ interactions model-independently by means of dispersion theory. We do not consider the $\eta\eta$ channel because, although the corresponding threshold is only about 100~MeV higher than the $K\bar K$ one, the latter is known to contribute significantly to
the $f_0(980)$ pole and couple strongly to the $\pi\pi$ channel. In contrast, the $\eta\eta$ channel is not essential for the $f_0$ states \cite{Kaiser:1998fi}, and therefore we expect its contribution to be marginal. Moreover, the effect of this channel
as well as other contributions like that from the $\omega$ exchange can be absorbed into the contact term included in the analysis (see below).

\section{Interaction between light hadrons and charmonia}
\label{sec:hhccint}

To get a quantitative understanding of the interaction of pions and vector charmonia, we may start from the transitions $\psi(2S)\to J/\psi\pi\pi$,
since excellent data exist for this process. The charged and neutral two-pion modes are the dominant decay modes of the $\psi(2S)$, with the branching fractions equal to $(34.68\pm0.30)\%$ and $(18.24\pm0.31)\%$, respectively~\cite{Zyla:2020zbs}. Note that transitions to a single
$\pi$ or $\eta$ are highly suppressed by the isospin and SU(3) symmetry, respectively.
Since the vector charmonia do not contain light quarks, the transition from the $\bar{c}c$
pair to the light mesons must be mediated by soft gluons under the approximation of the QCD multipole expansion~\cite{Gottfried:1977gp,Voloshin:1978hc}.  The underlying idea of the approach is that the size of a sufficiently heavy quarkonium $\bar{Q}Q$ is small compared to the inverse nonperturbative scale of QCD, $\Lambda_{\rm QCD}^{-1}$. This allows one to treat the heavy quarkonium as a compact object and consider the emission of soft (long-wave) gluons as a short-distance process. Then, at large distances, a well-developed method from the physics of atoms in the external electric field can be employed, and the interaction Hamiltonian can be written as
\be
H_{\rm int}\approx-\frac12 \xi_a \vec{r} \cdot \vec{E}^a,
\label{Hint}
\ee
where $\xi^a=t_1^a-t_2^a$ is the difference between the SU(3) color generators acting on the quark $Q$ and antiquark $\bar{Q}$, $\vec{r}$ is the relative position in the $\bar{Q}Q$ pair, and $\vec{E}^a$ is the chromoelectric field. Further calculations utilise an assumption that at a relatively long distance the gluons hadronise into light-quark hadrons such as pions, kaons, and so on, and effective theories for QCD at low energies can be further employed to evaluate the matrix elements of the gluonic operators between the asymptotic states formed by light hadrons. In particular, the amplitude of the dipion transition between two heavy quarkonia $A$ and $B$ takes the form
\be
{\cal M}(A\to B\pi\pi)=\alpha_{AB}\langle\pi\pi|\vec{E}^{a}\cdot \vec{E}^{a}|0\rangle,
\ee
with the effective coupling given by the so-called chromopolarisabilities defined as~\cite{Sibirtsev:2005ex}
\begin{equation}
\alpha_{AB}=\frac1{48}\langle B|\xi^a r_i G_O r_i \xi^a|A\rangle,
\label{alphaAB}
\end{equation}
where $G_O$ is the Green's function of the $\bar{Q}Q$ system in the color octet representation. Theoretical foundations of the gluon insertions in heavy-quark states were studied in Ref.~\cite{Peskin:1979va} where it is demonstrated that, for a sufficiently heavy meson, the interaction with external perturbations is described by the Wilson operator product expansion with the dependence on a particular heavy meson state encoded in the coefficients, that is, in particular, in the chromopolarisability (\ref{alphaAB}). Below we discuss a strong dependence of the chromopolarisability on the radial excitation number of the heavy $S$-wave quarkonia in the context of the effective interaction potential between two vector charmonia, in line with the findings of Ref.~\cite{Peskin:1979va}. For some recent discussions of the chromopolarisabilities within various approaches the interested reader is referred to, for example, Refs.~\cite{Brambilla:2015rqa,Eides:2015dtr, Eides:2017xnt,Polyakov:2018aey,TarrusCastella:2018php,Chen:2019hmz}. For a new formulation based on counting the dimension of the hybrid interpolating field, see Ref.~\cite{Pineda:2019mhw}.

\begin{figure}[t]
 \centering
 \includegraphics[width=\linewidth]{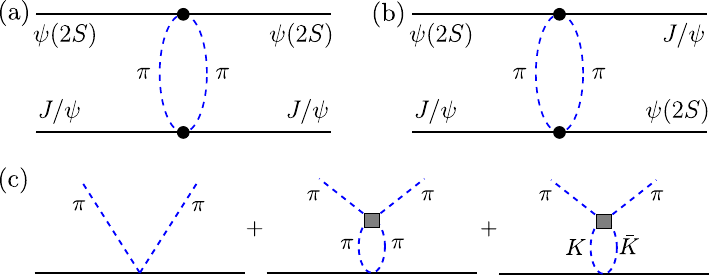}
 \caption{\co  Two-pion exchanges between the $\psi(2S)$ and $J/\psi$, where (a) is for $t$-channel and (b) is for $u$-channel. (c) The contributions to the amplitude for the transition $\psi_\alpha\to \psi_\beta\pi\pi$ including the coupled-channel ($\pi\pi$ and $K\bar K$) FSI. In (c), the solid and dashed lines label charmonia and light mesons, respectively, the filled squares denote the $\pi\pi$ and $K\bar K$ interactions and the black dots mean that the interactions between the light mesons, as depicted in (c),  have been taken into account.}
 \label{fig:feyn}
\end{figure}

Once the couplings of light meson pairs to the vector charmonia are known, one may evaluate the $\psi(2S)J/\psi$ scattering potential as depicted by diagrams in Fig.~\ref{fig:feyn}. The corresponding amplitude contains three different vertices and, therefore,
depends on three couplings, $\psi(2S)J/\psi\pi\pi$, $\psi(2S)\psi(2S)\pi\pi$, and $J/\psi J/\psi\pi\pi$, encoded in the corresponding chromopolarisabilities which satisfy the Schwarz inequality~\cite{Sibirtsev:2005ex},
\begin{equation}
\alpha_{J/\psi J/\psi} \alpha_{\psi(2S) \psi(2S)} \geq \alpha_{\psi(2S)J/\psi}^2.
\label{eq:inequality}
\end{equation}

Fig.~\ref{fig:feyn}b involves the polarisability $\alpha_{\psi(2S)J/\psi}$ from the right-hand side of Eq.~(\ref{eq:inequality}), which can
be extracted from the experimental data on the $\psi(2S)\to J/\psi\pi^+\pi^-$ decay, taking into account the $\pi\pi/K\bar K$ final state interaction (FSI). The formalism we employ is a straightforward extension of Refs.~\cite{Guo:2006ya,Chen:2019hmz} to a coupled channel interaction
 in the final state.

 Note that in order to connect the mentioned transitions to the polarisabilities we need to assume that contributions from hadronic loops in the
 $\pi$-vector charmonium channels can be neglected.\footnote{In Ref.~\cite{Du:2020bqj}, it was claimed that hadronic loops, which are non-multipole contributions, cannot be neglected
 when it comes to, for example, evaluating the $J/\psi$-nucleon
 interaction.}

\section{Coupled-channel FSI}
\label{sec:ccFSI}

The energy available for the dipion system in the decay $\psi(2S)\to J/\psi\pi\pi$ is approximately 600 MeV that implies that the FSI in the $\pi\pi$ system needs to be included, since in this mass range the $f_0(500)$ features prominently. Moreover, since
we intend to include the FSI in a model independent way by means of dispersion theory, the $\pi\pi$ phase shifts are needed as input over a significantly larger energy range. However, it is not clear what such $\pi\pi$ phase shifts, especially in the mass range above the two-kaon threshold, mean when the two-kaon channel is omitted. Therefore we extend the formalism
of Refs.~\cite{Guo:2006ya,Chen:2019hmz} to allow for the inclusion of the $K\bar K$ channel (see Fig.~\ref{fig:feyn} for the corresponding set of diagrams). To construct this amplitude we start from the chiral Lagrangian for the $\psi_\alpha\to\psi_\beta\Phi\Phi$ transition~\cite{Mannel:1995jt,Chen:2015jgl},
\begin{eqnarray}
\mathcal{L}_{\psi_\alpha \psi_\beta \Phi \Phi}&=&\frac{c_1^{(\alpha\beta)}}{2}\left\langle J^{\dagger}_\beta J_\alpha\right\rangle \Tr{u_{\mu} u^{\mu}}\nonumber\\[-2mm]
\label{eq:Lagrangian}\\[-2mm]
&+&\frac{c_2^{(\alpha\beta)}}{2}\left\langle J^{\dagger}_\beta J_\alpha\right\rangle\Tr{u_{\mu} u_{\nu}} v^{\mu} v^{\nu}+\text{H.c.}\nonumber,
\end{eqnarray}
where $\langle\,\cdots\rangle$ and Tr$[\cdots]$ mean the trace over spinor and flavour indices, respectively, $J_\alpha = \vec{\psi}_\alpha\cdot\vec{\sigma}+\eta_{c,\alpha}$, $\vec{\psi}_1$ ($\eta_{c,1}$) denotes the $J/\psi$ ($\eta_c$) charmonium field, $\vec{\psi}_2$ ($\eta_{c,2}$) denotes the $\psi(2S)$ ($\eta_c(2S)$), $\vec{\sigma}$ is the 3-vector made of Pauli matrices in the spinor space, $v_\mu$ is the four-velocity of the charmonia, and
\begin{align}
u_{\mu} &=i\left(u^{\dagger} \partial_{\mu} u-u \partial_{\mu} u^{\dagger}\right), \quad u=e^{i\Phi/(2F_\pi)}, \\
\Phi &=\left(\begin{array}{ccc}
\pi^{0} + \eta/\sqrt{3} & \sqrt{2} \pi^{+}&\sqrt{2}K^+ \\
\sqrt{2} \pi^{-} & -\pi^{0} + \eta/\sqrt{3}&\sqrt{2}K^0\\
\sqrt{2}K^-&\sqrt{2}\bar K^0& - 2\eta/\sqrt{3}
\end{array}\right), & \nonumber
\end{align}
where $F_\pi=92.2$~MeV is the pion decay constant.

Then the tree-level amplitude for the decay of $\psi(2S)$ into $J/\psi$ and an isoscalar meson pair reads~\cite{Chen:2015jgl,Chen:2016mjn,Baru:2020ywb}
\begin{eqnarray}
{\cal M}_{P}^{\rm tree}(s,\theta)&=&{\cal M}^S(s;m_P)+P_2(\cos\theta){\cal M}^D(s;m_P),\label{eq:M}
\end{eqnarray}
with the isoscalar $S$- and $D$-wave amplitudes,
\begin{eqnarray}
{\cal M}^S(s;m_P)&=&-\frac{2}{F_\pi^2}\sqrt{\frac{3}{2}} \left\{\vphantom{\frac12}c_1^{(21)}\left(s-2m_P^2\right)\right.\nonumber\\
&& +\left.\frac{c_2^{(21)}}{2}\left[s+\vec{k}^2\left(1-\frac{\sigma_P^2}{3}\right)\right]\right\}, \label{eq:M00}\\
{\cal M}^D(s;m_P)&=&\frac{2c_2^{(21)}}{3F_\pi^2}\sqrt{\frac{3}{2}}\vec{k}^2\sigma_P^2,\nonumber
\end{eqnarray}
respectively. Here the subindex $P= \pi (K)$ labels the contribution to the $\pi\pi (K\bar K)$ final state,
 $P_2(\cos\theta) = (3\cos^2\theta-1)/2$ is the second-order Legendre polynomial with the helicity angle $\theta$ being the angle between the moving direction of $\pi^+$ in the $\pi^+\pi^-$ center-of-mass (c.m.) frame and that of the $\pi^+\pi^-$ system in the $\psi(2S)$ rest frame, $\sigma_P=\sqrt{1-4m_P^2/s}$, and $\vec{k}$ is the 3-momentum of the $J/\psi$ in the $\psi(2S)$ rest frame.

 While both $S$ and $D$ waves are formally allowed in the $\pi\pi$ system, the decay $\psi(2S)\to J/\psi\pi^+\pi^-$ is dominated by the $S$-wave term, since a nontrivial contribution from the $D$ wave only appears above approximately 1~GeV due to the $f_2(1270)$ resonance, which lies beyond the phase space available in the studied decay. We, therefore, expect that the $D$ wave plays a minor role. Yet, it is necessary to reproduce the precisely measured angular distribution~\cite{Ablikim:2006bz}, which has a weak but nonvanishing angular dependence.

\begin{figure*}[thb]
 \centering
 \includegraphics[angle=270,width=\linewidth]{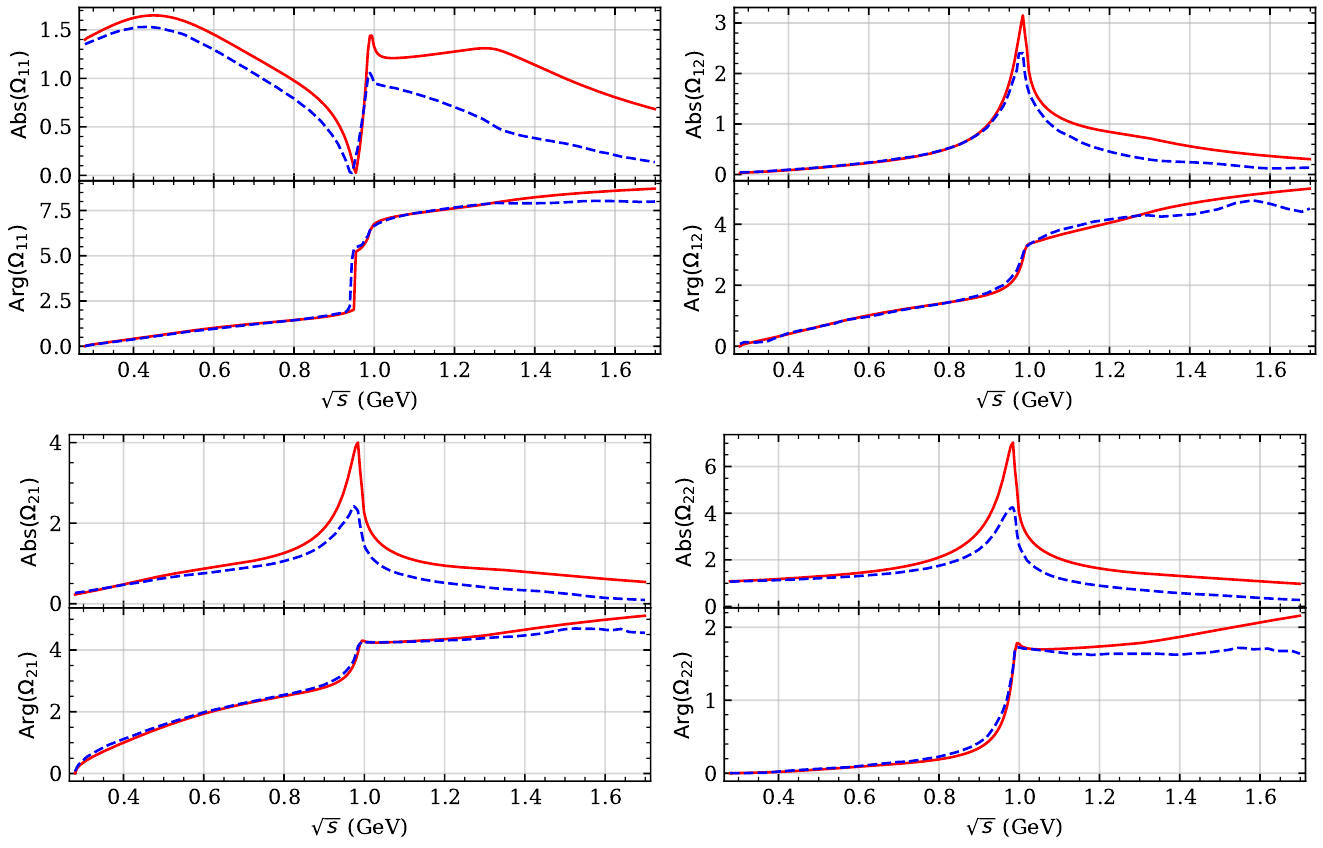}
 \caption{\co The absolute value and phase of the matrix elements of the Omn\`es matrix defined in Eq.~(\ref{Omnesmatrix}) taken from Ref.~\cite{Ropertz:2018stk} (red solid) and Ref.~\cite{Hoferichter:2012wf} (blue dashed).}
 \label{fig:Omnesmatrix}
\end{figure*}

For the $S$ wave, the coupled-channel FSI is introduced via the Omn\`es matrix,
\be
\hat{\Omega}^S(s)=
\left(
\begin{array}{cc}
\Omega_{11}^S(s) & \Omega_{12}^S(s)\\
\Omega_{21}^S(s) & \Omega_{22}^S(s)
\end{array}
\right),
\label{Omnesmatrix}
\ee
where the indices 1 and 2 stand for the $\pi\pi$ and $K\bar{K}$ channels, respectively (see Fig.~\ref{fig:Omnesmatrix} for the explicit form of the functions $\Omega_{ij}(s)$ ($i,j=1,2$) used in the calculations). The Omn\`es matrix elements taken from Ref.~\cite{Ropertz:2018stk}
were matched to the scattering phase shifts from Ref.~\cite{Dai:2014zta} at lower energies and, using the data
on $B_s\to J/\psi \pi\pi/K\bar K$, refined at higher energies above 1.3 GeV.

In the region above 1.3 GeV, where high precision determinations of the $\pi\pi$-$\bar KK$ parameters are not available and additional inelastic channels set in, different authors use different high-energy continuations which translate to polynomial differences in the Omn\`es matrix elements at lower energies. To illustrate the size of the difference, in Fig.~\ref{fig:Omnesmatrix},
we also show the matrix elements of Ref.~\cite{Hoferichter:2012wf} for comparison. In the energy range of relevance here (up to $\sqrt{s}= 0.6$ GeV --- see below), employing the different Omn\`es matrices leads to somewhat shifted values of the $c_i^{(21)}$ parameters. While this does not change significantly the extracted
 $J/\psi J/\psi$ scattering potential, where only the products
 of low energy constants and Omn\`es matrix elements enter and we anyway do not aim at high accuracy, we include this difference as theory uncertainty when extracting the chromoelectric polarisabilities below.

The total amplitude for the transitions to channel $i$ in $S$ wave which includes the $\pi\pi$-$K\bar{K}$ coupled-channel dynamics in the final state, as depicted in Fig.~\ref{fig:feyn}, reads~\cite{Ropertz:2018stk}
\begin{align}
\mathcal{M}_{i}(s){=}&\Omega^S_{i1}(s){\cal M}^S(s;m_\pi){+}\Omega^S_{i2}(s)\frac{2}{\sqrt{3}}{\cal M}^S(s;m_K),
\label{Mpipi}
\end{align}
where an isospin factor $2/\sqrt{3}$ is included for the kaonic amplitude~\cite{Chen:2015jgl,Chen:2016mjn,Baru:2020ywb}.

For the $D$ wave, we only consider the $\pi\pi$ FSI described by the Omn\`es function $\Omega^D(s)$, which is obtained from the $D$-wave phase shift, $\delta^D(s)$, via~\cite{Omnes:1958hv}
\be
\Omega^D(s)=\exp\left(\frac{s}{\pi}\int_{4m_\pi^2}^\infty\frac{\D x}{x}\frac{\delta^D(x)}{x-s}\right).
\ee
We use the parameterisation of the $D$-wave phase shift proposed in Ref.~\cite{GarciaMartin:2011cn} up to $\sqrt{s}=1.41$ GeV, which is extrapolated to $s\to \infty$ via
\begin{equation}
 \delta^D(s)=\frac{\pi}{2}+\arctan(a(s-b)),
\end{equation}
where $a$ and $b$ are two constants adjusted to match the phase shift at $\sqrt{s}=1.41$ GeV. Note that the large $s$ behaviour of $\delta^D$ is guided to $\pi$.

\begin{figure}[t]
 \centering
 \includegraphics[width=\linewidth]{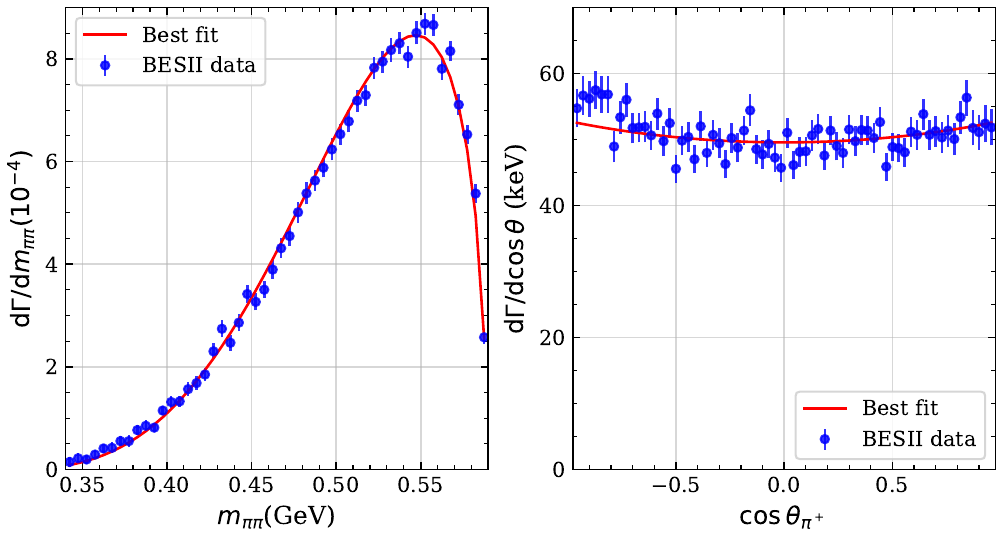}
 \caption{\co The fit to the BESII data on the $\psi(2S)\to J/\psi\pi^+\pi^-$ transition~\cite{Ablikim:2006bz} provided by Eq.~(\ref{Gammapipi}) with $\chi^2/$d.o.f. $=1.10$.}
 \label{fig:fit-dGam-cc}
\end{figure}

The resulting formula for the differential decay width ($m_{\pi\pi}=\sqrt{s}$),
\begin{equation}
\frac{\D\Gamma[\psi(2S)\to J/\psi\pi^+\pi^-]}{\D m_{\pi\pi}\D \cos\theta} = \frac{\sqrt{s} \sigma_\pi(s) |\vec k| }{128\pi^3 M_{\psi(2S)}^2 } \left| \mathcal{M}_{\pi\pi}(s,\theta) \right|^2,
\label{Gammapipi}
\end{equation}
with
\be
\mathcal{M}_{\pi\pi}(s,\theta)=\sqrt{\frac{2}{3}}\left[\mathcal{M}_{1}(s)+\Omega^D(s)P_2(\cos\theta)\mathcal{M}^D(s;m_\pi)\right],
\ee
contains two unknown parameters (low-energy constants (LECs) $c_1^{(21)}$ and $c_2^{(21)}$) and can be directly used to fit the BESII data on the dipion mass spectrum and angular distribution~\cite{Ablikim:2006bz} (see
Fig.~\ref{fig:fit-dGam-cc}). Employing the Omn\`es matrix of
Ref.~\cite{Ropertz:2018stk}, this gives the values
\be
c_1^{(21)}=0.229\pm0.002,\quad c_2^{(21)}=-0.061\pm0.003,
\label{c1c2}
\ee
which we use in the calculations below. Using the LECs $c_i^{(21)}$ as input, the absolute value for the transition polarisability $|\alpha_{\psi(2S)J/\psi}|$
can be calculated
(see, for example, Ref.~\cite{Chen:2019hmz} for the corresponding relations).\footnote{If the charmed meson loops play a sizeable role in the $\psi_\alpha \to \psi_\beta \Phi\Phi$ transitions, one can still use the chiral Lagrangian given in Eq.~\eqref{eq:Lagrangian}, and the effects can be absorbed into the coefficients $c_i^{(\alpha\beta)}$ as long as the involved charmonium masses are well below the thresholds of charmed hadron pairs. However, the relation between the $c_i^{(\alpha\beta)}$ values and the chromopolarisability will be lost.}
We find
\be
|\alpha_{\psi(2S)J/\psi}|= (1.81\pm 0.01\pm 0.11)~\mbox{GeV}^{-3},
\label{alpha12}
\ee
where the central value comes as an average of the polarisabilities extracted using two different Omn\`es matrices from Refs.~\cite{Ropertz:2018stk,Hoferichter:2012wf},
the first error is the statistical uncertainty from the fit and the second error is the uncertainty deduced from using two different Omn\`es matrices.
This value can be compared with the value $(1.66 \pm 0.01\pm 0.10)$~GeV$^{-3}$ obtained, if the coupled channel effects are neglected in the FSI and $(2.38 \pm 0.02)$~GeV$^{-3}$ if the FSI is turned off completely. It is reassuring that while the $K\bar K$ intermediate state gives some contribution, this contribution is significantly smaller than the one from the $\pi\pi$ interaction whose inclusion reduced the extracted polarisability from 2.4 to 1.7 GeV$^{-3}$, that is by
30\%. An additional inclusion of the $K\bar K$ intermediate state then enhances the polarisability by less than 10\%, which demonstrates a pleasing convergence in terms of the inclusion of increasingly heavy intermediate states.
The polarisability obtained in Ref.~\cite{Chen:2019hmz} is $(1.44\pm 0.02)$ GeV$^{-3}$, with only the $\pi\pi$ FSI included and with only the statistical uncertainty taken into account, which is within 2$\sigma$ of our result, when all coupled-channel effects are neglected. We traced this difference to the different FSIs employed, which supports that the difference in the Omn\`es functions leads to a significant uncertainty, in addition to the statistical one, which is, however, nicely captured by the uncertainties quoted above.

What is needed for the analysis of the $J/\psi J/\psi$ scattering potential are not
the $c_i^{(21)}$ parameters extracted from data above, but $c_i^{(11)}$. To account for possible differences between these quantities we introduce the parameters
\be
\xi_i \equiv {c_i^{(11)}}/{c_i^{(21)}}.
\label{xi}
\ee
Since both $c_1^{(\alpha\beta)}$ and $c_2^{(\alpha\beta)}$ are proportional to the corresponding chromopolarisibility~\cite{Chen:2019hmz}, we may write
\be
\xi_i\approx \xi \equiv \frac{\alpha_{J/\psi J/\psi}}{\alpha_{\psi(2S)J/\psi}}.
\label{xialpha}
\ee
Naively one may assume that $|\alpha_{J/\psi J/\psi}|\approx|\alpha_{\psi(2S)J/\psi}|$ which
translates to $|\xi|=1$. However, there are good
arguments in favour of larger values of $\xi$.

\begin{figure}
\centering
\includegraphics[width=0.8\linewidth]{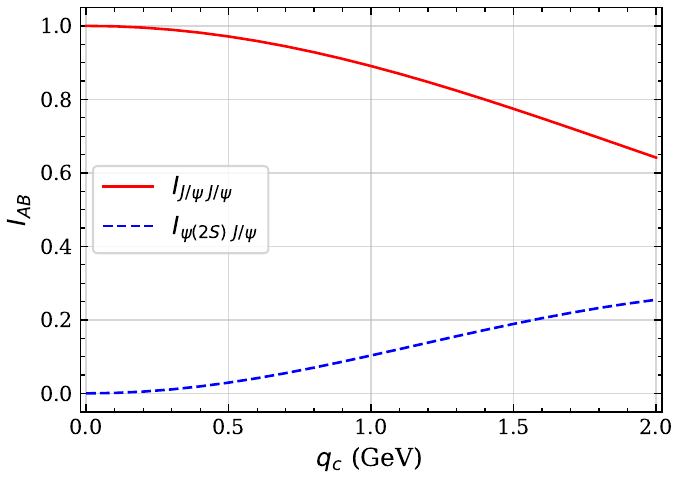}
\caption{\co The overlap integrals as defined in Eq.~(\ref{IAB}).}
\label{fig:IAB}
\end{figure}
At first, one may notice that a sizable contribution
to the polarisability $\alpha_{AB}$ should be proportional
to the overlap integral
\be
\alpha_{AB}\propto \int \mathrm{d}^3 r\; \psi_A^*(\vec{r})e^{-i\vec{q}_c\cdot\vec{r}/2}\psi_B(\vec{r})\equiv I_{AB}(q_c),
\label{IAB}
\ee
where $\psi_{A,B}(\vec{r})$ are the spatial wave functions of the initial- and final-state charmonia, respectively, which are determined by using the potential model in Ref.~\cite{Godfrey:1985xj}, while the operator $e^{-i\vec{q}_c\cdot\vec{r}/2}$ describes a shift in the $c$-quark momentum due to the emission of the two-pion system. The estimate in Eq.~\eqref{IAB} was obtained assuming that the emission of the two-pion system is driven by a one-body operator, which implies that the external current couples to a single charm (anti-)quark only. In Fig.~\ref{fig:IAB}, we plot the dependence of the overlap integrals $I_{J/\psi J/\psi}$ and $I_{\psi(2S) J/\psi}$ on the momentum transfer $q_c$ evaluated for realistic wave functions of the $J/\psi$ and $\psi(2S)$ charmonia. Naturally, $I_{J/\psi J/\psi}({q}_c=0)=1$ and $I_{\psi(2S) J/\psi}({q}_c=0)=0$, due to the orthonormality of
the wave functions. In addition,
the two functions also show a different behaviour
 as $q_c$ increases. For a more quantitative estimate
 of what these differences imply for the value of
 $\xi$,
we evaluate the ratio
\be
\frac{\int_{{q}_c<q_{\text{max}}}I_{J/\psi J/\psi}({q}_c)\mathrm{d}^3q_c}{\int_{{q}_c<q_{\text{max}}} I_{\psi(2S) J/\psi}({q}_c)\mathrm{d}^3q_c}\sim \mathcal O(10)\gg 1,
\label{Iratio}
\ee
with $q_{\rm max}\sim 2$ GeV. The value of this ratio
can be regarded as a very rough, order-of-magnitude estimate for the typical value of the ratios $\xi_i$ defined in Eq.~\eqref{xi}.

An alternative, and somewhat more rigorous, estimate of $\xi$ can be made with the help of the $J/\psi\pi$ scattering, whose interaction is weak enough to be treated perturbatively. The $J/\psi\pi$ scattering amplitude at threshold may be expressed in terms of the $S$-wave scattering length $a_0$ as
\begin{align}
{\cal M}_0^\text{th}[J/\psi\pi\to J/\psi\pi] =
{8\pi (M_{J/\psi}+m_\pi)}{a_0}.
\end{align}
The quantity $a_0$ has been calculated using lattice QCD~\cite{Yokokawa:2006td,Liu:2008rza} to be 0.0119(39) and $-0.01(1)$~fm, respectively (both results comply well with the theoretical estimate $|a_0|\lesssim 0.02$~fm from Ref.~\cite{Liu:2012dv}). Although these results look rather uncertain and even do not agree in the overall sign, they set a typical scale of this scattering length to be
\be
|a_0^{\text{lat}}|\sim 0.01~\text{fm}.
\label{alat}
\ee
To this end, we write the scattering length $a_0$ in terms of the low-energy constants $c_i^{(11)}$
and employ Eqs.~(\ref{eq:Lagrangian}), (\ref{c1c2}), (\ref{xi}) and \eqref{xialpha} to get
\be
a_0=\frac{
{ (c_1^{(11)}+c_2^{(11)}})m_\pi^2}{2\pi F_\pi^2 {(}M_{J/\psi}{ +m_\pi)}} \approx 0.0036\ \xi\ \rm fm,
\label{a}
\ee
and notice that relations (\ref{alat}) and (\ref{a}) can be reconciled with $\xi\simeq 3$ which appears to be in a reasonable agreement with the more naive estimate from Eq.~(\ref{Iratio}). Therefore, as a conservative estimate, we allow the ratios of the LECs
as defined in Eqs.~(\ref{xi}) and (\ref{xialpha}) to vary in the range
\be
1\lesssim \xi\lesssim 3
\label{alpharatio}
\ee
and then solve the Lippmann-Schwinger equation (LSE) for several representative values of $\xi$, including the two boundary cases.

\section{Interaction potential for $J/\psi J/\psi$ scattering}
\label{sec:JJint}

In this section, we derive the interaction potential between two $J/\psi$ mesons which arises from the correlated two-pion and two-kaon exchanges. The corresponding amplitude can be written as a sum
\be
\mathcal{M}_{J/\psi J/\psi}(s)=\mathcal{M}_\pi(s)+\mathcal{M}_K(s).
\label{MpiK}
\ee

The goal of this work is a study of the $J/\psi J/\psi$ scattering
very near threshold. Accordingly, we treat only this channel
dynamically and absorb all effects from higher channels
like $\psi(2S) J/\psi$, which were found to be important in Ref.~\cite{Dong:2020nwy}, into properly determined
contact terms (see discussion below). Then
 the interaction potential depends on
 the parameters $c_i^{(11)}$ introduced above and can be derived with the help of the dispersive formalism of Ref.~\cite{Donoghue:2006rg}. Employing unitarity, we write
\begin{eqnarray}
{\rm Im}\mathcal{M}_\pi(s)&=&|\mathcal{M}_1(s)|^2\rho_\pi(s),\nonumber\\[-2mm]
\label{ImMs}\\[-2mm]
{\rm Im}\mathcal{M}_K(s)&=&|\mathcal{M}_2(s)|^2\rho_K(s),\nonumber
\end{eqnarray}
with
\begin{align}
 \rho_\pi(s)&=\frac{\sigma_\pi(s)}{16\pi}
 \theta(\sqrt s-2m_\pi),\\
\rho_K(s) &= \frac{\sigma_K(s)}{16\pi}
\theta(\sqrt s-2m_K),
\end{align}
where $\theta(x)$ is the Heaviside step function. The amplitudes $\mathcal{M}_{1,2}$ are defined in Eq.~(\ref{Mpipi}) and, as before, the indices $i=1,2$ correspond to the $\pi\pi$ and $K\bar K$ channels, respectively.

Then the part of the interaction potential between two $J/\psi$
mediated by pairs of light pseudoscalar mesons reads
\begin{align}
V_{\text{exch}}(q)&=V_\pi(q)+V_K(q)\nonumber\\[-2mm]
\label{Vq2}\nonumber\\[-2mm]
&=-\frac{1}{4\pi M^2_{J/\psi}}\int_{4m_\pi^2}^\infty \mathrm{d}\mu^2\,\frac{{\rm Im}\mathcal{M}_{J/\psi J/\psi}(\mu^2)}{\mu^2+q^2},
\end{align}
where $q = |{\vec k}-{\vec k^\prime}|$ is the magnitude of the 3-momentum transferred between the initial and final meson pairs.

Let us investigate the convergence of the dispersive integral in Eq.~(\ref{Vq2}). The two-channel Omn\`es function behaves as $\propto 1/s$ as $s\to \infty$, so that $\mathcal{M}_{J/\psi J/\psi}(s)\sim\mathcal O(1)$. Therefore, the dispersive integral diverges logarithmically and needs regularisation. In what follows a Gaussian form factor $\exp[-(q^2+\mu^2)/\Lambda^2]$ is used to regularise it, so that the regularised potential reads
\begin{align}
 V_{\text{exch}}(q,\Lambda)=&\frac{-1}{4\pi M^2_{J/\psi}}\int_{4m_\pi^2}^\infty\mathrm{d}\mu^2\frac{{\rm Im}\mathcal{M}_{J/\psi J/\psi}(\mu^2)}{\mu^2+q^2}e^{-\frac{q^2+\mu^2}{\Lambda^2}}.
\label{Vq2reg}
\end{align}

This kind of form factor was proposed in
Ref.~\cite{Reinert:2017usi} in the context of a
high precision study of the nucleon-nucleon interaction within chiral effective field theory. The advantage of this choice is that for sufficiently large values of $\Lambda$ (of the order of the hard scale of the problem) the long-range parts of the potential remain
unchanged, when the regulator is introduced,
as we also demonstrate below for the system
studied here. This form factor at the same time
makes sure that no additional regulator is needed when the potential is introduced into
a LSE to calculate the full scattering amplitude, since the explicit
$q^2$ dependence leads to a sufficient suppression at large initial and final relative momenta.

\begin{figure}[t]
\centering
\includegraphics[width=\linewidth]{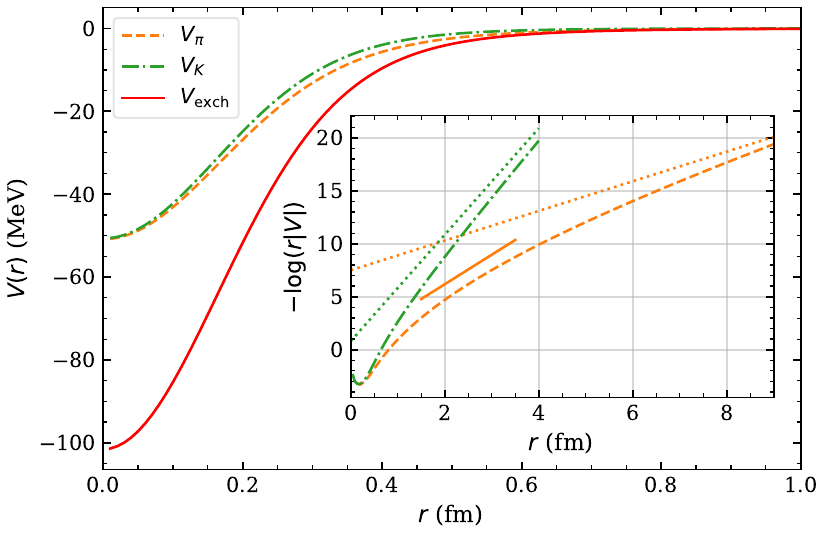}
\caption{\co The behaviour of the regularised potentials $V_\pi(r,\Lambda)$ and $V_K(r,\Lambda)$ as functions of $r$ for the cut-off $\Lambda=2$~GeV. The inlay demonstrates the mid- and long-range patterns for both potentials. To guide the eye, the dotted lines show the asymptotic slopes in the long-range:
$2m_\pi$ and $2m_K$ for the two-pion and two-kaon potential, respectively. The solid orange line in the inlay with a slope of $0.55$ GeV implies that the behaviour of the contribution from the $f_0(500)$ dominates in the mid-range (see the discussion in the text).}
\label{fig:Gau1}
\end{figure}

It is instructive to study the behaviour of the interaction potential in the coordinate space first. The Fourier transform of the potential (\ref{Vq2reg}) reads
\begin{align}
V_{\text{exch}}(r,\Lambda)=&-\frac{1}{4\pi M^2_{J/\psi}}\int\frac{\mathrm{d}^3q}{(2\pi)^3}e^{i\vec{q}\cdot\vec{r}}
\nonumber\\
&\times\int_{4m_\pi^2}^\infty\mathrm{d}\mu^2\,\frac{{\rm Im}\mathcal{M}_{J/\psi J/\psi}(\mu^2)}{\mu^2+q^2}e^{-\frac{q^2+\mu^2}{\Lambda^2}},\label{Vr}
\end{align}
and in Fig.~\ref{fig:Gau1} we demonstrate the behaviour of the two-pion and two-kaon components of this potential, $V_\pi(r,\Lambda)$ and $V_K(r,\Lambda)$, separately. The mid- and long-range patterns for both potentials in the logarithmic scale are shown in the inlay. To guide the eye, the dotted straight lines show the asymptotic slopes of $-\log(r|V|)$ for both potentials. For the given range in $r$, while the $2m_K$ asymptotic for the two-kaon exchange is quickly reached, the largest contribution to the two-pion exchange comes from the $f_0(500)$, so that the mid-range slope of the two-pion potential is close to 0.55 GeV~\cite{Donoghue:2006rg}.
Accordingly the two-pion exchange potential naturally approaches the limit $2m_\pi$ only in the long-range regime. Such a pattern can be easily understood if one looks at the behaviour of
the different elements of the Omn\`es matrix shown in
Fig.~\ref{fig:Omnesmatrix}: only $\Omega_{11}$ demonstrates a
very prominent structure from the $f_0(500)$ that accordingly
shows up only in the $\pi\pi$ exchange potential;
the other matrix elements are dominated by the $f_0(980)$
which, however, is located right at the $K\bar K$ threshold.
Therefore, the kaon potential approaches its asymptotic
slope ($2m_K$) without any delay.

We notice, however, that, since the potential (\ref{Vq2reg}) depends on the regulator, it is not well defined without a counter term (see, for example, Ref.~\cite{Baru:2015nea} for a similar discussion in the context of the one-pion exchange). We, therefore, augment it with a $\Lambda$-dependent contact term, $C$, regularised in a similar manner,
\be
V_{\rm CT}(q,\Lambda) = C \ e^{-\frac{q^2}{\Lambda^2}},
\ee
to arrive at the total potential in the form
\be
V_{\text{tot}}(q,\Lambda)=V_{\rm CT}(q,\Lambda)+V_{\text{exch}}(q,\Lambda),
\label{Vq2C}
\ee
where $V_{\text{exch}}(q,\Lambda)$ is defined in Eq.~(\ref{Vq2reg}) above. At the same time the contact term also absorbs
the effect from heavier scattering channels as well
as additional short-range contributions to the scattering potential.

\section{Strategy and results}
\label{sec:results}

The goal of the present study is to understand,
if the long and mid-range part of the $J/\psi J/\psi$ potential
derived above can provide a significant
contribution to the appearance of a
bound or virtual state near the $J/\psi J/\psi$ threshold. The $T$-matrix for the $J/\psi J/\psi$ scattering in the $S$ wave is calculated as a solution of the single-channel LSE,
\begin{eqnarray}\nonumber
T\left(E ; {k}^{\prime}, {k}\right)&=&V^S_{\text{tot}}\left({k}^{\prime}, {k},\Lambda\right)\\
& & {+}\int \frac{{\rm d}^{3} l}{(2 \pi)^{3}} \frac{V_{\text{tot}}^S\left({k}^{\prime},{l},\Lambda\right) T(E ; {l}, {k})}{E-l^{2}/M_{J/\psi}+i\epsilon},~~~
\label{LSEq}
\end{eqnarray}
where the $S$-wave interaction kernel $V^S_{\text{tot}}\left({k}^{\prime}, {k},\Lambda\right)$ is defined as
\be
V^S_{\text{tot}}\left({k}^{\prime}, {k},\Lambda\right)= \int \frac{d\Omega_{\vec{k'}}}{4\pi} V_{\text{tot}}(q,\Lambda),
\ee
with $V_{\text{tot}}(q,\Lambda)$ from Eq.~(\ref{Vq2C}).
Note that the integral equation is well defined even without additional regulator,
since the way we regularise the two-meson exchange integrals automatically
regularises the integral in the LSE (see the discussion below Eq.~(\ref{Vq2reg})). For the renormalisation
procedure we adjust the strength of the contact
term such
 that the $T$-matrix has a pole either 1 or 5~MeV below the double-$J/\psi$ threshold,  that is in the energy range consistent with the analysis of Ref.~\cite{Dong:2020nwy}, on the physical or
unphysical sheet referring to a bound or
virtual state, respectively. This way the
regulator dependence gets absorbed into the contact
term and the $T$-matrix is, at least close
to the pole, regulator-independent as indicated in the notation of Eq.~(\ref{LSEq}). The renormalisation
implies necessarily that the relative importance of the contact and exchange potentials depends on the cut-off (see, for example, Ref.~\cite{Epelbaum:2003xx}).  Accordingly, without an additional input, all conclusions about the role of the $\pi\pi / K\bar K$ exchanges can only be qualitative but not quantitative. To still address
the physics issue raised above, we
perform the calculations for
\be
1~\mbox{GeV}\leq\Lambda\leq {2}~\mbox{GeV}.
\ee
Within this interval we regard values
of $\Lambda\leq 1.5$ GeV as phenomenologically adequate as we have included both the $f_0(500)$ and $f_0(980)$ states explicitly via the dispersion formalism and neglected higher resonances.
 Then for different values of the parameter $\xi$, introduced in Eq.~(\ref{alpharatio}), we
study if the mentioned virtual or bound states emerge under the condition

\be
R\equiv\frac{V^S_{\text{exch}}(k^\prime=0,k=0,\Lambda)}{V^S_{\text{tot}}(k^\prime=0,k=0,\Lambda)}\gtrsim 1/2 \, ,
\label{VexVtot}
\ee
{\B because this range of the ratio} would indicate that a significant fraction of the double-$J/\psi$ binding
potential comes from the exchange of
pairs of light mesons that we largely control quantitatively.
{\B Indeed, since the interaction between two charmonia is suppressed by either the Okubo-Zweig-Iizuka rule (exchange of soft gluons or light mesons at the hadronic level) or by $\Lambda_\text{QCD}^2/m_c^2$ (exchange of charmonia), there is a priori no reason for the contact term to be much larger  than the two-pion and two-kaon exchanges. Thus, from the consideration of naturalness, they are expected to be of the same order of magnitude.} 

\begin{figure}[t]
\centering
\includegraphics[width=\linewidth]{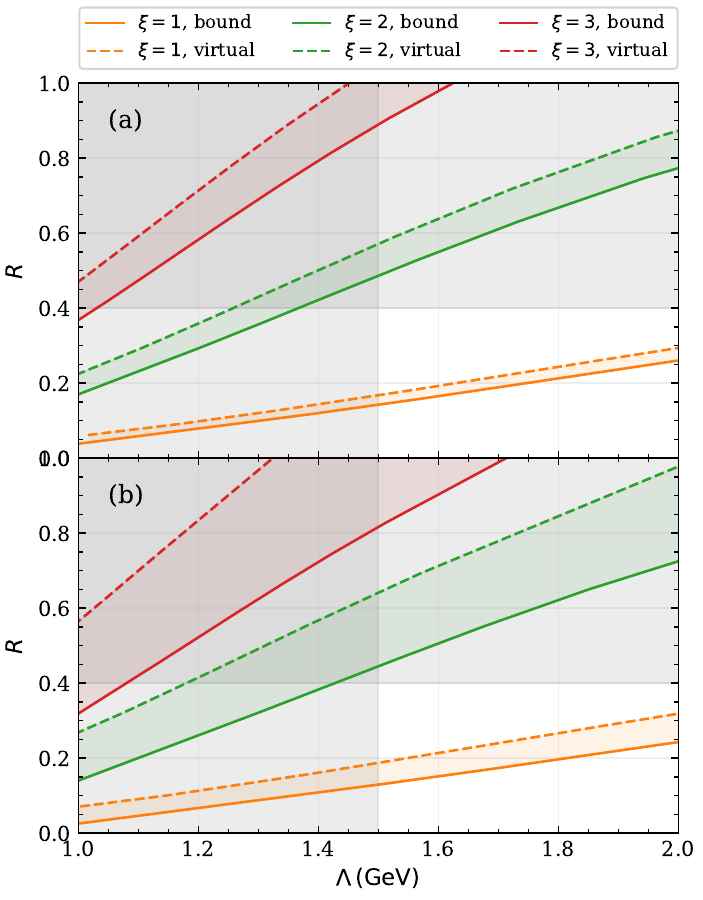}
\caption{\co The dependence of the ratio
$R$ on the cut-off $\Lambda$ {\B for  two fixed locations of the generated pole, namely $E_{\rm pole}=1$ MeV (upper panel) and $E_{\rm pole}=5$ MeV (lower panel) below the double-$J/\psi$ threshold used as input} for a bound state (solid line) and   a virtual state (dashed line).
The ratio $\xi$ defined in Eq.~(\ref{xi}) is fixed to take three representative values compatible with Eq.~(\ref{alpharatio}). Notice that $\xi$ is very likely larger than 1. The vertical and horizontal bands indicate the range of the cut-offs and the relative contribution from the exchange of pion and kaon pairs, respectively, considered as reasonable in this study. {\B To guide the eye, as lower bound for the
allowed range of the ratio of the potentials 0.4 is chosen which is consistent with Eq.~(\ref{VexVtot}).
The intersection area of the two bands in the upper left corner of each plot may be considered as the most natural
parameter range for the emergence of a near-threshold state largely driven by the $\pi\pi /K\bar K$ dynamics.}
}
\label{fig:VEpercent}
\end{figure}

\begin{figure}[t]
\centering
\includegraphics[width=\linewidth]{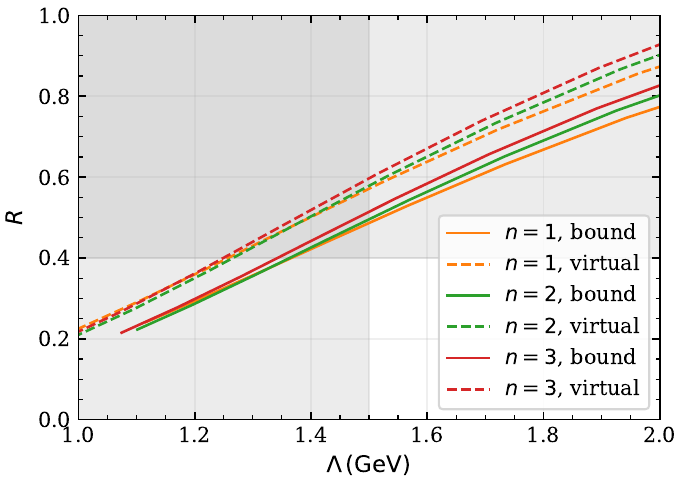}
\caption{\co Sensitivity of the results for  the ratio $R$ to the shape of the regulator. We use a generalised Gaussian form factor $\exp[-(q^2+\mu^2)^n/\Lambda^{2n}]$ to regularise the meson exchanges and $\exp[-q^{2n}/\Lambda^{2n}]$ for the contact term with $n$=1, 2, and 3. The results are illustrated for $\xi=2$ and $E_B=1$ MeV and demonstrate a very weak dependence on $n$. For a description of the shaded area, see the caption of Fig.~\ref{fig:VEpercent}.
}
\label{fig:VEpercent2}
\end{figure}

The results of our calculation are shown in Fig.~\ref{fig:VEpercent}. The phenomenologically adequate ranges for the cut-off and the ratio (\ref{VexVtot}) are highlighted by the vertical and horizontal shaded bands, respectively. {\B We would also like to stress once again that a particular choice of the boundary value of the ratio (\ref{VexVtot}) is somewhat arbitrary and should only be considered as a qualitative estimate. Nevertheless the message conveyed by Fig.~\ref{fig:VEpercent} is quite clear. Namely,
}
one can see that already values of $\xi$ from the middle of the allowed interval (\ref{alpharatio}) are consistent with the existence of both a bound and virtual state in the double-$J/\psi$ system. The results only weakly depend on the shape of the regulator used (see Fig.~\ref{fig:VEpercent2}). We interpret these findings such that the emergence of a $J/\psi J/\psi$ near-threshold bound state is very natural and consistent with our present understanding of the meson-exchange forces.

\section{Summary and outlook}
\label{sec:summary}

In this work, we have studied the interaction in a double-$J/\psi$ system mediated by the correlated
$\pi\pi$ and $K\bar K$ exchanges. The primary transitions between
vector charmonia emitting the light hadrons are induced by soft gluon exchanges that can be constrained by
experimental data as well as theoretical
considerations.
{ Given a lack of the experimental information on the $J/\psi J/\psi$ system, we start from an educated guess that the meson-exchange picture should be important for binding if the $\pi\pi / K\bar K$ meson-exchange potential $V_{\rm exch}$ for small momenta is at least roughly about one half of the full potential $V_{\rm tot}$, where the latter can also contain the contact interaction from other short-range sources, the origin of which is obscure. Then, we demonstrate that this ratio can indeed be achieved using plausible values for the ratio of the chromopolarizabilities $\xi $ and for reasonably low cut-offs. To illustrate this explicitly, in Fig. 7, we set the binding energy for the $J/\psi J/\psi$ bound (or virtual) state to be 1~MeV, which is consistent with the analysis of Ref.~\cite{Dong:2020nwy}.\footnote{ This binding energy defines the long-range tail of the wave function in the double-$J/\psi$ system and, therefore, the mean separation between the constituents can be estimated as $1/\sqrt{m_{J/\psi} E_B}\sim 3.5$~fm which is a fairly large distance, sufficient for the hadronisation of gluons that is an important prerequisite for the analysis reported in this work.}
This allows us to adjust the contact term for any given cut-off in a way consistent with renormalisation group invariance and, therefore, to compare $V_{\rm exch}$ with $V_{\rm tot}$ directly. }
We found that, although the $J/\psi\pi$ interaction is rather weak with the scattering length of only $\mathcal{O}(0.01~{\rm fm})$, the proposed two-meson exchange potential already gives a significant
contribution to a possible binding of
a near-threshold state. The existence of
such a state finds evidence in a description of the recent LHCb data of the $J/\psi J/\psi$ production within a coupled-channel approach of Ref.~\cite{Dong:2020nwy}.
Given very conservative restrictions imposed on all parameters of the model, we regard this result as robust.
{ A qualitative support of our findings comes from the recent lattice QCD result about the existence of a near-threshold bound state of $\Omega_{ccc}\Omega_{ccc}$ when the Coulomb repulsion is switched off~\cite{Lyu:2021qsh}, which should have similar strong interaction as a pair of charmonia.}
A direct measurement of this near-threshold double-$J/\psi$ state, for example, by studying the $J/\psi e^+e^-$ or $J/\psi \mu^+\mu^-$ system would be highly desired.

If the polarisability $|\alpha_{\psi(2S)\psi(2S)}|$ is assumed to be of the same order of magnitude as $|\alpha_{J/\psi J/\psi}|$, then a similar pole near the double-$\psi(2S)$ threshold may also be expected. The existence of a sibling pole at the $\psi(2S)J/\psi$ threshold would also be possible considering the diagrams in Fig.~\ref{fig:feyn}.
However, the coupling of these systems to the lower double-$J/\psi$ channel requires a more complicated coupled-channel analysis which lies beyond the scope of the present research.

As a possible complementary research we suggest that the $P$-wave $J/\psi\pi$ scattering be studied in lattice QCD because in this case, unlike the $S$-wave scattering as given in Eq.~(\ref{a}), the low-energy limit of the $P$-wave scattering amplitude,
\be
{\cal M}_1[J/\psi\pi\to J/\psi\pi] \approx
 8\pi (M_{J/\psi}+m_{\pi})
k^2 a_1,
\ee
is sensitive not to a particular combination of the short-range constants $c_1^{(11)}$ and $c_2^{(11)}$ but basically to only one of them,
\be
a_1=-\frac{M_{J/\psi}c_1-m_\pi c_2}{6\pi F_\pi^2 M_{J/\psi}(M_{J/\psi}+m_\pi)}\approx -\frac{c_1}{6\pi F_\pi^2 M_{J/\psi}}.
\label{a1}
\ee
From the considerations above we expect the constant $a_1$ to be, approximately,
\be
a_1\simeq -(0.2\sim0.6)~\text{GeV}^{-3}.
\ee
A lattice calculation of both $a_0$ and $a_1$ will allow one to fix the mid- and long-range potential between two $J/\psi$ mesons using the formalism in this work. Such a calculation is free of any disconnected Wick contractions as well as of the complications due to coupled channels. Once the results are available, one can proceed to more robust predictions for the possible existence of a near-threshold state in the double-$J/\psi$ system.

\section*{Conflict of interest}
The authors declare that they have no conflict of interest.

\section*{Acknowledgments}

This work was supported in part by the Chinese Academy of Sciences (CAS) under Grants No. XDPB15, No. XDB34030000, and No. QYZDB-SSW-SYS013, by the National Natural Science Foundation of China (NSFC) under Grants No. 11835015, No. 12047503 and No. 11961141012, and by the NSFC and the Deutsche Forschungsgemeinschaft (DFG) through the funds provided to the Sino-German Collaborative Research Center ``Symmetries and the Emergence of Structure in QCD'' (NSFC Grant No.~12070131001, DFG Project-ID 196253076 -- TRR110). The work of A. N. was supported by Ministry of Science and Education of Russian Federation under Grant 14.W03.31.0026. X.-K. D. is grateful for the valuable discussions with Meng-Lin Du, Wei Hao, Hao-Jie Jing and Zhen-Hua Zhang. He also thanks Yun-Hua Chen for providing the BESII data on $\psi(2S)\to J/\psi \pi^+\pi^-$ and communications regarding the details of the calculation in Ref.~\cite{Chen:2019hmz}.

\section*{Author contributions}

Xiang-Kun Dong and Feng-Kun Guo did the calculations. Alexey Nefediev and Christoph Hanhart drafted the manuscript. All the authors made substantial contributions to the physical and technical discussions and the editing of the manuscript. All the authors have read and approved the final version of the manuscript.

\end{document}